# Investigating the use of ChatGPT for the scheduling of construction projects

**Samuel A. Prieto [1], Eyob T. Mengiste [1], Borja García de Soto [1]**

[1] S.M.A.R.T. Construction Research Group, Division of Engineering, New York University
Abu Dhabi (NYUAD), Experimental Research Building, Saadiyat Island, P.O. Box 129188,
Abu Dhabi, United Arab Emirates
samuel.prieto@nyu.edu, eyob.mengiste@nyu.edu, garcia.de.soto@nyu.edu

**Abstract**
Large language models such as ChatGPT have the potential to revolutionize the construction industry by automating repetitive and time-consuming tasks. This paper presents a study in which ChatGPT was used to generate a construction schedule for a simple construction project. The output from ChatGPT was evaluated by a pool of participants that provided feedback regarding their overall interaction experience and the quality of the output. The results show that ChatGPT can generate a coherent schedule that follows a logical approach to fulfill the requirements of the scope indicated. The participants had an overall positive interaction experience and indicated the great potential of such a tool to automate many preliminary and time-consuming tasks. However, the technology still has limitations, and further development is needed before it can be widely adopted in the industry. Overall, this study highlights the potential of using large language models in the construction industry and the need for further research.

## 1 Introduction

Natural Language Processing (NLP) combines areas such as linguistics, computer science, and Artificial Intelligence (AI) and focuses on the interaction between computers and humans using programs that are developed from large natural language data [1]. Selected applications of NLP in the construction industry include (1) Extracting information from construction documents: NLP techniques can extract relevant information, such as specifications, plans, and contracts, and convert it into a structured format that can be quickly processed by computers [2]. This can help streamline the process of reviewing and comparing construction documents and reduce the risk of errors caused by manual data entry. (2) Analyzing construction site data: NLP can be used to analyze construction site data, such as progress reports, safety inspections, and quality control reports, and extract insights that can help improve project efficiency and mitigate risks [3]. Improving communication on construction sites: NLP-powered chatbots and virtual assistants can be used to improve communication on construction sites by providing a quick and convenient way for project stakeholders to access information and ask questions about the project [4]. (3) Generating or enhancing construction schedules: With enough information about the project being provided to the system, NLP techniques can be used to generate construction schedules based on project details provided by a user. This is not new in the construction field, and there are plenty of tools well developed that generate optimized construction schedules. However, the way this information is presented to the user is always schematized and structured in charts, tables, and diagrams. A natural language discussion could offer a new point of view in this regard [5].

In summary, NLP has the potential to significantly improve efficiency, accuracy, and overall communication in the construction industry. Within the field of NLP, research has been focused on developing Large Language representation Models (LLM) that expand the applicability of NLP to more detailed human language understanding tasks such as translation, text classification, and holding conversations [6,7].

## 1.1 Aims and contribution

In this paper, we evaluate the applicability of a Generative Pre-Training Transformer language representation model (GPT) to assist in developing an automated construction schedule based on natural language prompts. The aims of this study are summarized as follows:

- Explore the possible applications and limitations of a rapidly growing and powerful tool for construction scheduling and resource loading.
- Conduct a preliminary case study involving multiple users applying GPT to generate a resource-loaded project schedule for a simple project based on a given detailed natural language description input (i.e., prompt).
- Evaluate the results obtained from the participants in the case study based on parameters such as accuracy, efficiency, clarity, coherence, reliability, relevance, consistency, scalability, and adaptability.

The rest of the paper is organized as follows: Section 2 provides a brief state-of-the-art study on various language representation models and the current trend in project management tools. Section 3 presents the methodology followed to evaluate the results from the case study. Section 4 presents the case study. Section 5 discusses the results and findings from the case study, going into detail about the strengths and limitations of the tool. Finally, Section 6 contains some conclusions and future work by the authors on the application.

## 2 Literature review

### 2.1 Language representation models

Large language models have gained significant attention in recent years due to their ability to generate human-like text and perform a wide range of language-based tasks. In the construction industry, large language models have the potential to improve efficiency, accuracy, and communication in several different ways. The fields where NLP technology has been tested and applied are limited. The application of NLP technologies in the construction sector is limited. Some examples include the work by Xue et al. [2] that used NLP techniques to summarize construction contracts. Locatelli et al. [8] studied the potential of combining NLP and BIM, focusing on automated compliance checking and semantic BIM enrichment goals.

Regarding language representation models, different approaches have been developed in the past decade [7]. The major ones can be grouped into three families: the autoregressive language model Generative Pre-trained Transformer (GPT), the Bidirectional Encoder Representations from Transformers (BERT) and the Multi-Task Learning.

### 2.1.1 Generative Pre-trained Transformer (GPT)

The GPT family (Generative Pre-trained Transformer) of language models was developed by OpenAI. The models are trained on a large dataset of text and, as expected, are able to generate human-like text. The most recent version of the GPT family is GPT-3 [9], which is the base for ChatGPT (considered as GPT-3.5), the tool used for this paper. GPT-3, released in June 2020, has 175 billion parameters, making it one of the largest language models to date. Since its release, it has been proved that GPT-3 is capable of performing some tasks that traditionally require human-level understanding, such as writing essays and programming. Despite the technology not being particularly new [10], GPT-3.5 has been fine-tuned for information retention during the conversation, making it suitable for the scope being tested in this paper. This feature has motivated researchers to study the possibility of incorporating such language models into activities that were solely reserved for human-human interaction, such as healthcare delivery [11]. Floridi and Chiriatti [12] studied the scope, limits and consequences of the GPT-3 model and how society will have to get used to not being able to tell if a text was written by a human or an AI.

### 2.1.2 Bidirectional Encoder Representations from Transformers (BERT)

BERT (Bidirectional Encoder Representations from Transformers) is a pre-trained transformer-based neural network model developed for NLP tasks, such as text classification and question answering (chatbots). BERT was first introduced by Google in 2018 [13] and has been the state-of-the-art for many of the NLP models developed afterward. Hassan et al. [14] used a BERT-Based model to identify risky and hazardous situations related to construction. Moon et al. [15] used a BERT model for the automated detection of contractual risk clauses from construction specifications. Their approach classified contractual risk categories to provide reviewers with crucial clauses that commonly cause disputes, such as payment, temporal, procedure, safety, role and responsibility, definition and reference. Yao and Garcia de Soto [16] used a BERT model classification for semantic screening trained on construction-specific documentation to investigate main topics related to construction cybersecurity.

### 2.1.3 Text-to-Text Transfer Transformer (T5) Multi-tasking learning

T5 (Text-to-Text Transfer Transformer) is a multi-task pre-training model for NLP tasks. T5 was introduced by Google in 2020 [17]. T5 models are trained to perform multiple tasks at once, which allows them to learn a general-purpose understanding of language that can be fine-tuned for specific tasks.

## 2.2 Automation of construction scheduling

The area of schedule automation has been heavily researched in the past few decades. Modern construction project scheduling can generally be categorized as BIM Driven and Machine Learning based schedule generations.

### 2.2.1 BIM-driven schedule generation

BIM models comprise geometric information (special relationships of components), materials and resources. Under the current industrial practice, project management teams utilize the information from the BIM to optimize and generate schedules. Researchers improved the process of manual schedule development by automating task dependency and sequencing using pre-set rules, patterns or pre-set knowledge learned from historic cases[18,19].

These methods extract required data inputs from the pre-built information model to produce a schedule with a logical order [20]. However, BIM models often do not include environmental factors, temporary structures, equipment and material availability, specific construction specifications and methodology of the building site. Therefore, BIM-based automatic generation of schedules requires manual intervention to be practical [20,21].

### 2.2.2 Machine Learning based schedule generation

Several approaches have been proposed to perform predictive modeling to utilize historical data to predict project outcomes [22]. However, these approaches generally rely on large amounts of data to train the models. The source data could be visual [23], where the algorithm understands the characteristics of the sites and work performance.

Language-based methods were also the focus of research for automated construction schedule development. Natural language models such as GPT [5] or Language clustering methods such as Latent Dirichlet Allocation (LDA) and Latent Semantic Analysis (LSA) [24] were used to understand human language of explaining tasks develop dependency and clustering.

Recently, Hong et al. [20] proposed a graph-based construction scheduling approach. Their proposed approach stores past best practices and recycles the information to optimize resource usage, task sequence and duration of the new project.

In general, Machine Leaning based approaches produced promising results for construction schedules;

however, the performance of most methods depends on the availability of data and the data processing technical and infrastructural capacity of the user.

## 3 Methodology

A small experiment has been designed to evaluate the potential of ChatGPT as an aid in project management in construction, focusing on project scheduling and task assignment. The experiment consists of a simple construction project (scope) in an existing space. The goal is to retrieve a logical and accurate task breakdown from ChatGPT. The same experiment was performed by different participants, allowing them to challenge the tool and modify the original plan to evaluate the ChatGPT response.

Various parameters were used to evaluate the results of these experiments, including accuracy, efficiency, clarity, coherence, reliability, relevance, consistency, scalability, and adaptability. Accuracy was measured by comparing the ChatGPT-generated project schedules and task assignments to a baseline schedule and assignments created by a human project manager. Efficiency was evaluated by collecting data about the time required to create a schedule and assign tasks using ChatGPT, as well as the time needed to fix the number of errors or mistakes made throughout the process. Clarity and relevance were assessed by evaluating the proposed schedules and responses subject to modifications and complexity changes to ensure they were clear and easy to understand. The responses must be relevant to the input prompt and not deviate from the original task. Coherence was measured by examining the results, such as task dependencies or crew assignments, to ensure that they made sense from a logical point of view in the eye of an expert. Reliability was evaluated by checking conformity with standards and assessing its adequateness if those results were to be used in a real project. Consistency was measured by evaluating the invariance of the ChatGPT responses based on slight modifications to the initial prompt, or multiple instances of the same prompt. The scalability and adaptability of the ChatGPT solution were also evaluated, including its ability to handle larger projects or deal with additional tasks or responsibilities as needed and its ability to handle changing project requirements or unforeseen challenges.

The initial input for the study consisted of a paragraph with enough details of the scope to be completed (see Section 4). Details included the description of the work to be done, dimensions and the type of material to be used, the expected level of completeness of the task, and the due date for the task to be completed. The results expected from the process were a list of tasks and overall scheduling (sequence) of the project, with the ability to interactively react to changes made by the user. An overview of the methodology is shown in Figure 1.

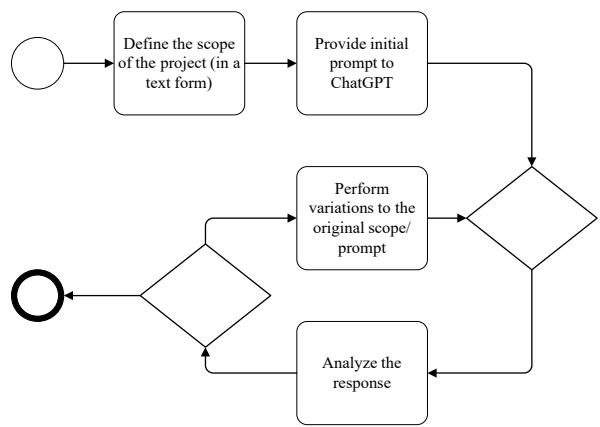

Figure 1. Overview of the main steps to use and assess ChatGPT in this study.

The original scope underwent small modifications concerning the original prompt to challenge ChatGPT. Some modifications included adding a new scope (e.g., electrical or plumbing work).

A survey was conducted containing instructions to set the bases of the experiment and a series of questions to evaluate the quality of the output generated by ChatGPT and the participants' experience interacting with it. The results are summarized in Section 5.

## 4 Case study

A simple project and scope were used to evaluate the performance of ChatGPT when providing a construction schedule. In general, the work consisted of the addition of a partition wall in an existing space. A simplified floorplan is shown in Figure 2. To ensure that all the participants provided the same information regarding the scope, an initial prompt was given to ChatGPT in all cases. The initial prompt was:

*"A set of instructions on a construction project will be provided. You will store the provided information, and you won't provide any answers to the initial prompt until asked otherwise."*

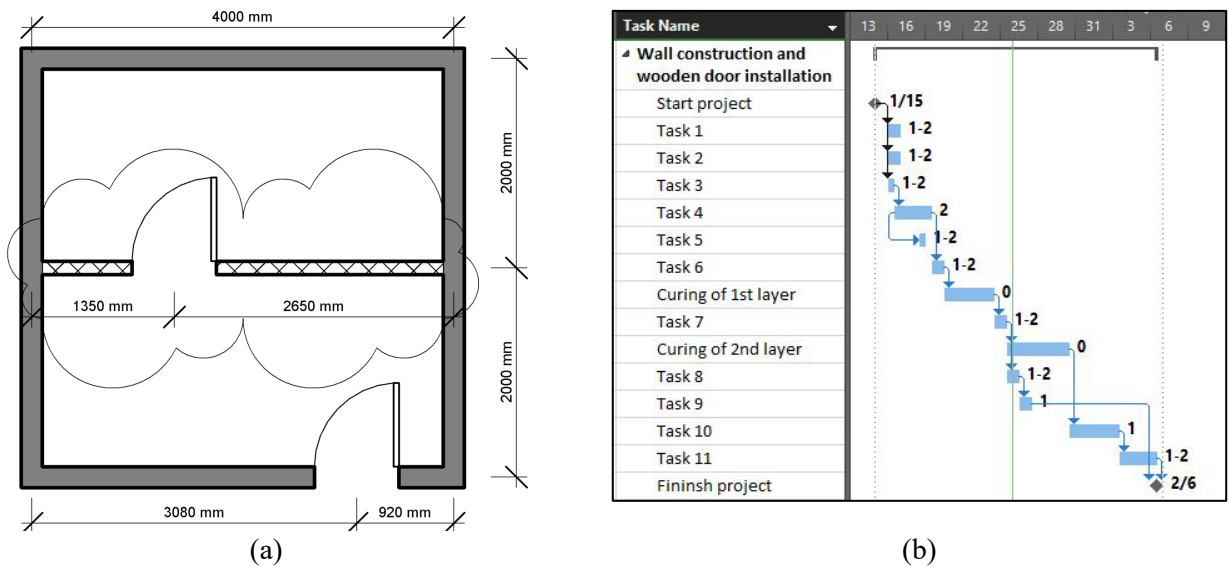

Figure 2. (a) General view of the floorplan and (b) Gantt chart of main activities used as the baseline for the required scope summarized in Table 1.

The initial input describing the scope was:

*"A new partition needs to be done in an already existent space, where the new partition is grouted with the existing walls. The details of the room to be partitioned are the following: the room is rectangular shaped, 4 meters by 4 meters in total. The walls are made of concrete masonry units. The height of the walls is 3 meters, and the width is 20 centimeters. The new partition needs to be made out of concrete masonry units as well. The partition is meant to split the original space in half, resulting in two individual spaces of approximately 4 meters by 2 meters. The partition needs to account for the installation of a single solid, two-panel wooden door of 0.8 meters in width by 2.1 meters in height and 35 mm thickness that will communicate the two new spaces. After the partition is made, it needs to be plastered with two layers of stucco and painted with two layers of white latex paint on both sides of the wall. No electrical or plumbing installation is needed. No ceiling work is needed. The floor is cement screed. The work needs to be completed in less than three weeks."*

A typical schedule for the proposed scope is shown in Table 1. This was used as the baseline for comparison with the output generated by ChatGPT. Based on the initial prompt, ChatGPT was asked to use that information to generate a construction schedule to complete the scope. To do that, the following prompt was asked to ChatGPT:

*"Can you come out with a suitable project schedule?"*

To structure the obtained information into usable data, the following structure was asked from ChatGPT:

*"Based on the details of the work to be completed, extract the information in the following structure 'task name / task priority / task dependencies / number of people needed / expected duration of task'"*

The prompt *"Add this information into columns"* could be used to show the output in a table format. Based on this simple scope, a series of questions with a set of initial instructions were put in a survey form. This form was distributed to several individuals (participants) with different skill sets and qualifications working in the construction and AI fields. The feedback from the six participants is summarized in Section 5.

Table 1. Data considered as the baseline for the performed case study.

| Task No. | Task name | Task dependencies | People needed* | Expected duration |
|---|---|---|---|---|
| 1 | Inspect the existing space and check proposed work is in line with existing conditions | - | 1-2 | 1 day |
| 2 | Prepare the work area and protect surrounding areas as needed | 1 | 1-2 | 1 |
| 3 | Measure and mark the location of the new partition, including the location for opening (door) | 1 | 1 | 1 |
| 4 | Install CMU for new partition | 1 | 2 | 3 |
| 5 | Install framing for the new door | 4 | 1-2 | 1 |
| 6 | Apply the first stucco layer to the CMU wall – includes curing time | 4 SS+2 | 1-2 | 3 |
| 7 | Apply the second stucco layer to the CMU wall – includes curing time | 6 | 1-2 | 4 |
| 8 | Install and adjust the wooden door | 7 | 1 | 0.5 |
| 9 | Protect the door in preparation for the painting of the new CMU partition wall | 8 | 1 | 0.5 |
| 10 | Finish wall (prime, paint, apply 2 layers - allow drying time per manufacturer's recommendations) | 7 | 2 | 4 |
| 11 | Clean up and final inspection | 10 | 1 | 1 |
| | | | TOTAL | 15.5 days** |

*RS Means was considered when estimating the people needed
**Total duration, including weekends (equivalent to 11.5 work days)

## 5 Results and discussion

The entire output regarding the schedule obtained by the 6 different participants is summarized in Table A.1. The table contains all the information regarding the different tasks proposed by ChatGPT, their dependencies, the assigned priority, the expected number of people and the time needed to complete them. In general, ChatGPT provided a logical (although very linear) sequence of tasks in all cases. The tool could extrapolate a breakdown of the steps needed without that information being explicitly provided and establish logical and coherent dependences amongst the proposed tasks. At first glance, the output seemed coherent and reasonable, and the participants were awed by the speed at which the responses were provided (in most cases, within a few seconds of entering the prompt). However, with further inspection, it was clear that not all proposed tasks agreed with the scope of work.

Table 2 shows a direct comparison of the tasks proposed by ChatGPT in all six cases with respect to the baseline. It is worth noticing that the tasks related to the wooden door (i.e., the frame installation and its protection before painting) were not considered in any of the schedules proposed by ChatGPT. This is due to the fact that ChatGPT has not been trained for specific construction purposes, and it is not aware that for the door installation, the frame needs to be placed first. In some cases, the two layers of plastering are proposed as one single task, and most of the planning and preparation are joined into a single task. In addition, three of the six responses by ChatGPT included the demolition of the existing wall, which is not relevant to the scope provided. This was probably wrongly inferred by ChatGPT based on the information about a *"new partition needs to be done in an already existent space,"* and demolition might have been assumed because of the existing space condition. Also, incorrect information was provided regarding the tasks that would be expected. For example, in three of the six cases, ChatGPT indicated the installation of a foundation, which might not have been required for a partition wall. In one case, it also suggested the installation of steel rebar for the new partition and the placement ("pouring") of concrete for the new partition. However, minor schedule errors can be fixed by conversing with ChatGPT, instructing the tool to rearrange some of the tasks, asking for a more detailed breakdown, or providing more information that was not properly understood, such as the frame installation for the door.

Table 2. Comparison of the tasks proposed in the schedules generated by ChatGPT with the ones from the baseline.

| Task no. | Task name (Baseline) | Participant 1 | 2 | 3 | 4 | 5 | 6 |
|---|---|---|---|---|---|---|---|
| 1 | Inspect the existing space and check proposed work is in line with existing conditions | ✗ | ✗ | ✗ | ✗ | ✓ | ✓ |
| 2 | Prepare the work area and protect surrounding areas as needed | ✗ | ✗ | ✓ | ✗ | ✓ | ✓ |
| 3 | Measure and mark the location of the new partition, including the location for opening (door) | ✗ | ✗ | ✓ | ✗ | ✓ | ✓ |
| 4 | Install CMU for new partition | ✗ | ✓ | ✓ | ✓ | ✓ | ✓ |
| 5 | Install framing for the new door | ✗ | ✗ | ✗ | ✗ | ✗ | ✗ |
| 6 | Apply the first stucco layer to the CMU wall – includes curing time | ✓ | ✓ | ✓ | ✓ | ✓ | ✓ |
| 7 | Apply the second stucco layer to the CMU wall – includes curing time | ✓ | ✓ | ✓ | ✓ | ✓ | ✓ |
| 8 | Install and adjust the wooden door | ✓ | ✓ | ✓ | ✓ | ✓ | ✓ |
| 9 | Protect the door in preparation for the painting of the new CMU partition wall | ✗ | ✗ | ✗ | ✗ | ✗ | ✗ |
| 10 | Finish wall (prime, paint, apply 2 layers - allow drying time per manufacturer's recommendations) | ✓ | ✓ | ✓ | ✓ | ✓ | ✓ |
| 11 | Clean up and final inspection | ✓ | ✓ | ✓ | ✓ | ✓ | ✓ |

The proposed dependencies and sequence of tasks are logical for the most part, despite being very linear. The proposed sequence for installing the wooden door is arguably not ideal. ChatGPT includes the door installation right after the wall erection. In general, the sequence to install the door would be after the plastering, as indicated in the baseline schedule, to avoid damage from intermediate tasks. If only the door frame is considered, the suggested sequence would be adequate; however, a clear breakdown for that was not provided or inferred by ChatGPT.

In order to evaluate both the duration and crew needed for each task, only those tasks present in the baseline were considered. The durations for each task reported by ChatGPT are summarized in Table A.1. The variation between the baseline durations (Table 1) and the ones from ChatGPT are displayed in Figure 3. The bars represent the difference between the proposed time by ChatGPT in each of the different survey participants' experiments and the one present in the baseline. The bars without data are for tasks that were not considered by ChatGPT. For a given task, a positive increment indicates that the duration from ChatGPT is greater than the one from the baseline and vice versa. Values equal to 0 indicate that the duration from ChatGPT is the same as the one in the baseline. For example, for Task 4, participant 1 (P1) did not have this task, so the value for that is empty. For P2 and P6, the duration obtained from ChatGPT was the same as the baseline (i.e., the deviation is 0). For P3, the duration from ChatGPT was one day more than the baseline (3 days in the baseline vs. 4 days in ChatGPT). For P4 and P5, the duration from ChatGPT was one day less than the baseline (3 days in the baseline vs. 2 days in ChatGPT). Some of the tasks with the biggest difference are those that involve drying time (i.e., Tasks 6, 7, and 10). In the baseline, drying time has been considered both for the plastering and the painting, which is why those tasks take more than three days each. However, ChatGPT does not seem to be considering drying times.

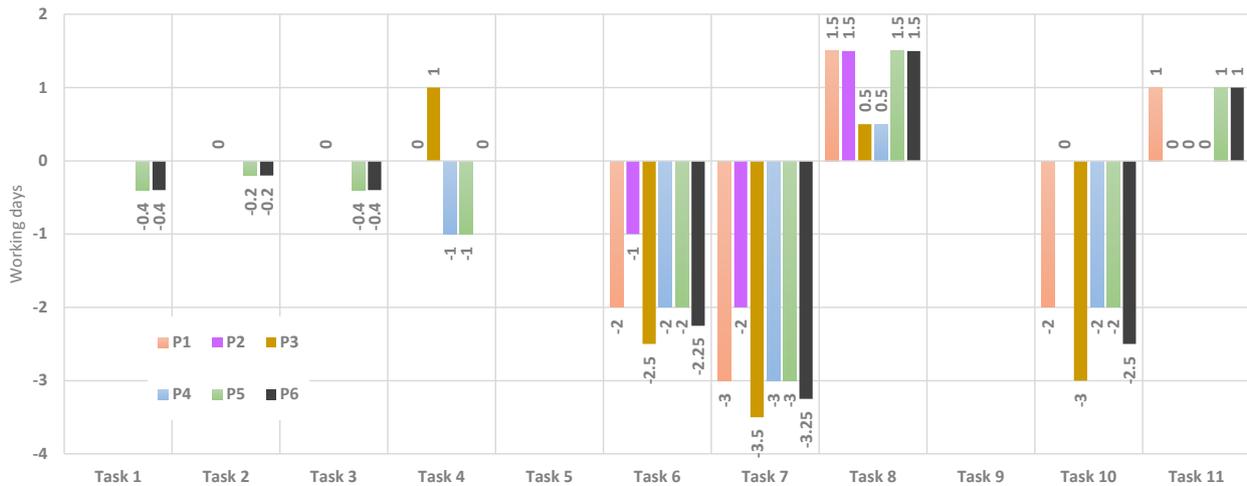

Figure 3. Deviation between the durations proposed by ChatGPT and the baseline for each task from each participant.

The estimation of the number of workers reported by the participants using ChatGPT was compared with the baseline estimation. Most of the number of workers per task from all the ChatGPT responses and the baseline are given as a range of maximum and minimum numbers. Therefore, the comparison was done considering the maximum and the minimum possible difference in workers per task per participant. The highest deviation from the baseline estimation is two workers. The results are shown in Figure 4. Positive increments indicate an increase in the number of workers compared to the baseline and vice versa. For instance, the estimated number of workers for Task 6 in the baseline ranged from 1 to 2. P1 and P4 reported that the ChatGPT estimated the same amount. However, for P3, ChatGPT estimated that Task 6 would require 2 workers. Therefore, based on the difference between the baseline (considering the range) and this response, ChatGPT estimated that Task 6 would require a maximum of 1 more worker (in the case of the lowest baseline estimate) and a minimum of zero additional workers (in the case of a maximum estimate) as represented by the bar for P3 in Task 6. Similarly, P5 and P6 estimated that a minimum of 2 and a maximum of 3 workers would be required to complete task 6. This estimate is one worker higher than the baseline estimate, both in the case of upper and lower baseline estimates.

In order to test the adaptability and scalability of ChatGPT, the original prompt was modified to add additional components to the scope, such as electrical or plumbing work. For example:

*"The new wall that needs to be built for the partition needs to have a couple of electrical sockets. Can you include that task in your schedule and regenerate it?"*

ChatGPT was able to modify the original list of tasks and update the sequence accordingly. An example of the new task for the example above is shown in Table 3.

To further challenge ChatGPT, information regarding the cost was also asked. The following prompt was used:

*"Based on the information above, can you provide a cost estimate for labor and materials needed?"*

As expected, the responses from ChatGPT were vague and very generic, constantly advising to consult with a licensed contractor to get a rough cost estimation. A sample response from ChatGPT was, *"I'm sorry, but I am not able to provide you with a cost estimate for labor and materials needed based on the information* *provided. The cost of construction projects can vary greatly depending on a variety of factors such as location, materials, labor costs, and the specific requirements of the project. Additionally, factors such as the cost of permits, inspections, and any necessary equipment rentals should also be considered. It would be best to consult with a licensed contractor or estimator to provide you with an accurate cost estimate for your project."* When insisting on providing a rough estimate, the response was still generic (e.g., *"As a language model AI, I don't have access to current market pricing and cost may vary greatly depending on location, materials, and labor. However, for a rough estimate for a partition wall and door installation in a standard room, you could expect to spend anywhere from $1000 to $5000 or more for a standard partition wall and door installation. This is a rough estimate and the real cost could vary greatly depending on the location, the specific needs of the project, and the contractor you choose. It is important to consult with a local licensed contractor or estimator to provide you with an accurate cost estimate for your project."*) This shows that one of the big limitations of the current model is that the data is not updated since it is not connected to real-time internet data. Since the model is taught to be aware of its own limitations, the default response from ChatGPT is to rely on a licensed contractor for such a query.

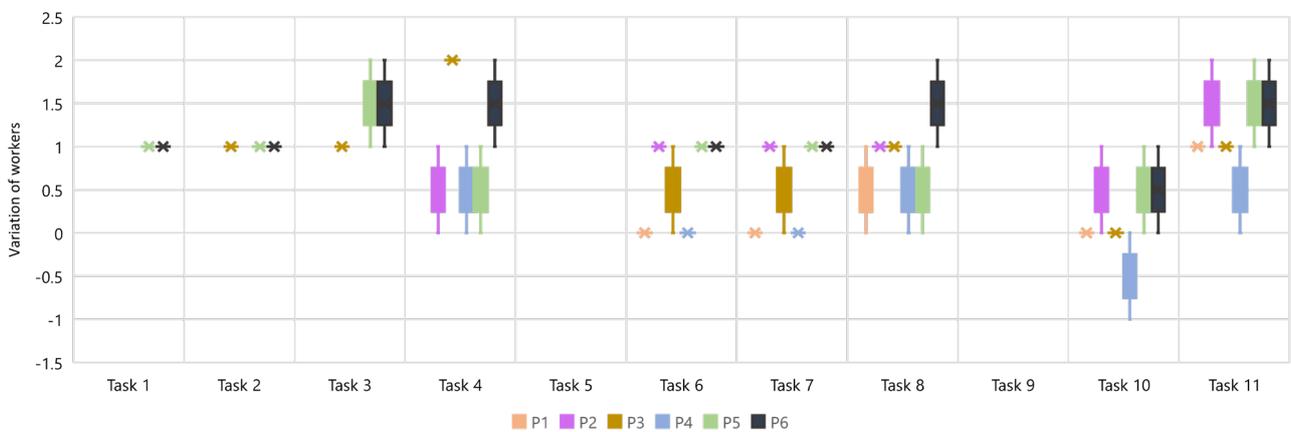

Figure 4. Deviation of the needed workers proposed by ChatGPT with respect to the baseline.

Regarding the quality of the output generated by ChatGPT, the participants' general impression was positive. They were impressed by the fact that ChatGPT could produce a (for the most part) logical and coherent task breakdown with little initial information, despite being a model not trained specifically for construction purposes. The results are summarized in Figure 5.

Table 3. Information regarding the installation of a couple of electrical sockets.

| Task name | Dependencies | People needed | Expected duration |
|---|---|---|---|
| Install electrical sockets | Building the new partition | 2 | 4 hours |

The overall evaluation is very positive, with ratings of "good" and "very good" predominating in all the different features. Accuracy and reliability got the lowest scores due to the issues related to using tasks that were not related to the scope of work (e.g., excavation, foundation work, rebar, etc.) when compared to the baseline, therefore making the results not as reliable as they would need to be trusted in a professional environment.

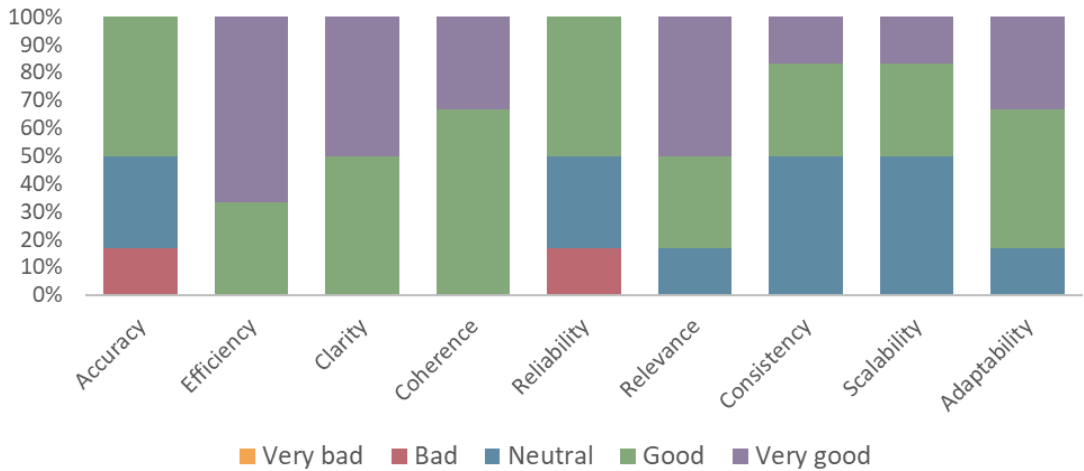

Figure 5. Results from the survey regarding the evaluation of the ChatGPT output.

In addition to the evaluation of the performance of ChatGPT regarding the proposed tasks and sequence (i.e., scheduling component), a set of qualitative questions were asked to evaluate the overall impressions of the participants regarding their experience interacting with ChatGPT. This included their impression of the type of communication over the classical methods (i.e., charts and tables) and their feedback about potential uses for ChatGPT in the construction field. The results regarding the interaction with ChatGPT are presented in Figure 6. Overall, the interaction was very positive, with most participants rating different aspects of the interaction "very good", such as how intuitive it was, how comfortable it was, how efficient it was, and the overall interaction experience.

Regarding the preference for this communication (i.e., in a dialog format) instead of the classical methods (lists, tables, charts), all the surveyed participants shared the common opinion on tables and charts being more intuitive and easier to read than the results provided by ChatGPT. Nonetheless, they all agreed that despite this not being the preferred final form for displaying the results, it might become an important and useful tool to extract the information to be displayed.

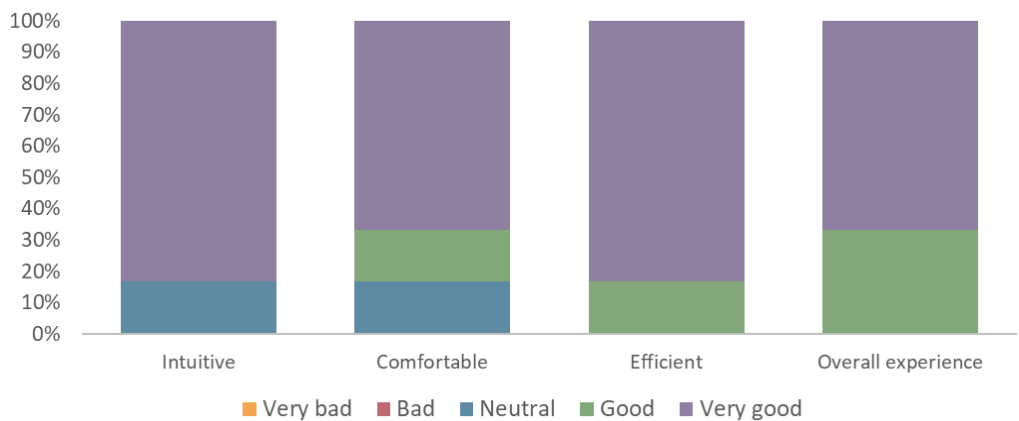

Figure 6. Results from the survey regarding the evaluation of the interaction with ChatGPT.

The possible uses proposed by the surveyed participants ranged from initial consulting, safety and risk identification, basic design, cost estimation, processing and evaluation of contracts or identifying vulnerabilities.

Overall, they all agreed that any task involving text processing could benefit from the inclusion of Natural Language Processing techniques. The full data regarding the performed survey, with all the responses from the participants and their full conversation with ChatGPT, can be made available to interested readers upon reasonable request to the authors.

## 5.1 Limitations

The presented case study has been useful for preliminary testing and evaluating the possible uses and capabilities of ChatGPT when applied to the generation of a schedule for simple projects. The results are promising; however, the complexity of the case study is very limited. Further studies considering increased complexity to resemble actual construction projects should be conducted to further consider this technology in the construction field.

The results provided are based on a limited number of participants and should not be used for generalization purposes. A bigger statistical pool of surveyed people is needed for a wider generalization of the results obtained.

## 6 Conclusions and future work

Since ChatGPT has been made available, it has attracted researchers from different fields. The application of such tools in the construction industry should not be overlooked. This study consisted of a simple example to assess the applicability of such a tool in the context of project scheduling. The performance and results were very promising for the simple use case, especially considering that ChatGPT has not been specifically trained for such an application. However, several significant flaws would limit the application of such a tool in a real project. Having said that, the overall performance was reasonable, and the interaction experience was positive. This is important to show that such technology could be relatively easy to integrate if consistent and reliable performance is achieved. Such tools that are specialized in a given field (e.g., project scheduling) could become extremely beneficial and play an important role in the automation of repetitive and time-consuming parts.

Future research needs to be conducted to further explore the applicability and capabilities of Natural Language Processing tools in the construction industry. To do that, a GPT model will be specifically trained for construction purposes and challenged with a more complex scenario. A larger pool of participants will be surveyed in a future study to more accurately generalize the results. In addition, the way the initial information is given to ChatGPT could become a nuisance for larger and more complex projects. This is why it is worth exploring new methods to input information into these models, such as in the form of floorplans and images.

**Acknowledgment**

The authors thank the participants who provided their feedback and experience using ChatGPT for the case study.

**Appendix A**

Table A.1. Data extracted from the first experiment by the six participants.

| Participant | Task name | Dependencies | Priority* | People needed | Expected duration |
|---|---|---|---|---|---|
| 1 | 1. Demolition of existing wall | - | H | 2-3 | 1 day |
| | 2. Framing of partition | 1 | H | 2-3 | 2 days |
| | 3. Measurement and cutting of wooden door | 2 | M | 1-2 | 1 day |
| | 4. Installation of wooden door | 3 | M | 1-2 | 1 day |
| | 5. Plastering of partition | 4 | L | 1-2 | 2 days |
| | 6. Final touch-ups and cleanup | 5 | L | 2 | 2 days |
| | 7. Painting of partition | 5 | L | 2 | 2 days |
| | | | | TOTAL | 11 days |
| 2 | 1. Demolition of existing wall | - | H | 2-3 | 2 days |
| | 2. Lay the foundation for the new partition | 1 | H | 2-3 | 3 days |
| | 3. Build the partition wall up to the ceiling | 2 | H | 2-3 | 3 days |
| | 4. Install the wooden door | 3 | M | 2 | 2 days |
| | 5. Apply the first layer of stucco | 4 | M | 2-3 | 2 days |
| | 6. Apply the second layer of stucco | 5 | M | 2-3 | 2 days |
| | 7. Apply the first layer of paint | 6 | M | 2-3 | 2 days |
| | 8. Apply the second layer of paint | 7 | M | 2-3 | 2 days |
| | 9. Final inspection and cleaning up | 8 | H | 2-3 | 1 day |
| | | | | TOTAL | 19 days |
| 3 | 1. Prepare the site | - | H | 2-3 | 1 day |
| | 2. Measure and mark the existing space | 1 | H | 2 | 1 day |
| | 2. Excavation for the new partition's foundation | 2 | H | 2-3 | 2 days |
| | 4. Building the new partition | 3 | H | 4 | 4 days |
| | 5. Install the wooden door | 4 | H | 2 | 1 day |
| | 6. Apply two layers of stucco to both sides of the partition | 3 | H | 2 | 1 day |
| | 7. Apply two layers of white latex paint to both sides of the partition | 6 | H | 2 | 1 day |
| | 8. Clean up the site | 7 | H | 2 | 1 day |
| | | | | TOTAL | 12 days |
| 4 | 1. Demolition of existing wall | - | H | 2-3 | 1 day |
| | 2. Excavation and preparation of foundation for new partition | 1 | H | 2-3 | 1 day |
| | 3. Installation of steel reinforcement for new partition | 2 | H | 2-3 | 1 day |
| | 4. Pouring of concrete for new partition | 3 | H | 2-3 | 1 day |
| | 5. Erection of new partition using concrete masonry units | 4 | H | 2-3 | 2 days |
| | 6. Installation of wooden door | 5 | H | 1-2 | 1 day |
| | 7. Plastering of new partition with first layer of stucco | 5 | H | 1-2 | 1 day |
| | 8. Plastering of new partition with second layer of stucco | 7 | H | 1-2 | 1 day |
| | 9. Painting of new partition with first layer of white latex paint | 8 | H | 1-2 | 1 day |
| | 10. Painting of new partition with second layer of white latex paint | 9 | H | 1-2 | 1 day |
| | 11. Clean up and final inspection | 10 | H | 1-2 | 1 day |
| | | | | TOTAL | 13 days |

* H=High, M=Medium, L=Low

Table A.1. (continued)

| Participant | Task name | Dependencies | Priority* | People needed | Expected duration |
|---|---|---|---|---|---|
| 5 | 1. Preparation and planning | - | H | 2-3 | 2 days |
| | 2. Demolition of existing walls | 1 | H | 2-3 | 2 days |
| | 3. Construction of new partition | 2 | H | 2-3 | 2 days |
| | 4. Installation of wooden door | 3 | H | 1-2 | 2 days |
| | 5. Plastering | 3 | H | 2-3 | 2 days |
| | 6. Painting | 5 | H | 2-3 | 2 days |
| | 7. Clean up and final inspection | 6 | H | 2-3 | 2 days |
| | | | | TOTAL | 14 days |
| 6 | 1. Preparation and planning | - | H | 2-3 | 2 days |
| | 2. Demolition | 1 | H | 2-3 | 3 days |
| | 3. Masonry work | 2 | H | 3-4 | 3 days |
| | 4. Door installation | 3 | H | 2-3 | 2 days |
| | 5. Plastering and painting | 4 | M | 2-3 | 3 days |
| | 6. Cleanup and final inspection | 5 | L | 2-3 | 2 days |
| | | | | TOTAL | 15 days |

* H=High, M=Medium, L=Low